\documentstyle[twocolumn,epsf]{jpsj}
\def\runtitle{Superconductivity of Q1 and Q2 Dimensional
Tight-Binding Electrons in Magnetic Field}

\title{Superconductivity of Quasi-One and Quasi-Two Dimensional
Tight-Binding Electrons in Magnetic Field}
\author{Mitake {\sc Miyazaki},  Keita {\sc Kishigi} and
 Yasumasa {\sc Hasegawa}}
\inst{Faculty of Science, Himeji Institute of Technology,\\Kamigouri-cho, Akou-gun, Hyogo 678-1297, Japan }
\recdate{\hspace{4cm}    }
\abst{The upper critical field $H_{c2}(T)$ of the tight-binding
electrons in the three-dimensional lattice is investigated.
 The electrons make Cooper pairs between the eigenstates with the same
energy in the strong magnetic field. The transition lines in the  
quasi-one dimensional case are shown
to deviate from the previously obtained results where the 
hopping matrix
elements along the magnetic field are neglected. In the absence of the
Pauli pair breaking the transition temperature $T_c(H)$ of the quasi-two
dimensional electrons is obtained to oscillationally increase as the
magnetic field becomes large and reaches to $T_c(0)$ in the strong field
as in the quasi-one dimensional case.}
\kword{field-induced superconductivity, organic conductors, 
quasi-two-dimension, quasi-one-dimension}
\begin{document}
\sloppy
\maketitle

In the semiclassical approximation, the upper critical field $H_{c2}(T)$ 
of isotropic superconductors is derived from
Ginzburg-Landau-Abrikosov-Gor'kov (GLAG) theory.~\cite{rf:1,rf:2} 
Lawrence and Doniach~\cite{rf:3} have applied GLAG theory to layered
superconductors. Klemm {\it et al}.~\cite{rf:4} and Bulaevskii and
Guseinov.~\cite{rf:5} have shown that
$H_{c2}(T)$ displays upward curvature when the magnetic field is applied
parallel to the layer. When the coherence length $\xi(T)$
of the order parameter
 becomes smaller than the distance of layers, $H_{c2}(T)$ is
infinite.
 However, these results are based on a semiclassical approximation, in
which only the phase of the wave function of electrons is changed by
 magnetic field.

It was shown that the quantum effect of the electrons in the BCS theory lead
to reentrant behavior in high magnetic field.
~\cite{rf:6,rf:7,rf:8,rf:9,rf:10,rf:11} The reentrance of the
superconductivity due to Landau quantization 
 has attracted theoretical interest.~\cite{rf:7,rf:8}
 In that case, Cooper pairs are formed between 
electrons at the lowest Landau level in strong magnetic field. However,
when only the lowest Landau level is filled in the three dimensional
case, the system can be treated as a 1D system, i.e., energy depends
only on the momentum parallel to the magnetic field, and it has been
shown that the system is unstable to the density wave state rather than
the superconductivity.~\cite{rf:12,rf:13}

Recently, superconductivity in strong magnetic field is observed  
in organic superconductors (TMTSF)$\mbox{}_2$PF$\mbox{}_6$.
~\cite{rf:14,rf:15} Similar result has also been observed in
(TMTSF)$\mbox{}_2$ClO$\mbox{}_4$.~\cite{rf:15,rf:16} Organic 
superconductors (TMTSF)$\mbox{}_2$X are well described by quarter-filled
tight-binding electrons (actually holes) 
with $t_a\sim 3000$K, $t_b/t_a\sim 0.1$, $t_c/t_a\sim 0.003$.
Since the magnetic field of 10T corresponds to $\phi/\phi_0\sim 1/1000$
in (TMTSF)$\mbox{}_2$X, where $\phi$ is flux per unit area and $\phi_0$
is the flux quantum, the effect of the Landau level quantization is
negligible. However, when the magnetic field is applied perpendicular to
plane, field-induced spin density wave (FISDW) is stabilized. The
existence of FISDW shows that the quantum effect is important in
quasi-one dimensional (Q1D) systems and the semiclassical approximation
of the magnetic field is not appropriate in these systems.

Lebed~\cite{rf:6} has predicted that the Q1D superconductors should 
exhibit superconductivity in a strong magnetic field. The
superconductivity observed in (TMTSF)$\mbox{}_2$PF$\mbox{}_6$ and 
(TMTSF)$\mbox{}_2$ClO$\mbox{}_4$ is thought to be the realization of the
Q1D superconductivity in strong magnetic field.
The mechanism of the reentrance of Q1D superconductivity is
similar to that of FISDW. In
the presence of the magnetic field, the dimensionality of the system is
reduced. When the magnetic field is applied in the $c$ direction in the
system with hopping matrix elements $t_a\gg t_b\gg t_c$, the effect of
$t_b$, which makes the nesting of the Fermi surface imperfect with the
imperfectness parameter of the order of $t_b^2/t_a$, disappears and
spin-density-wave is induced. When the magnetic field is applied in $b$
direction, the nesting of the Fermi surface stays imperfect but the
orbital frustration is removed if we take account of the eigenstates in
the magnetic field. In the Q1D case, the magnetic field necessary for the
reentrance of the superconductivity is much smaller than that in the
case of Landau level quantization. 
Dupuis {\it et al}.~\cite{rf:9} have extensively studied
 the mean-field transition line $T_c(H)$ of 
 Q1D superconductors and shown the cascade transitions to the
superconducting states. We have studied the
anisotropic superconductivity with line nodes of the energy gap,
~\cite{rf:10} which is
thought to be realized in organic superconductors
(TMTSF)$\mbox{}_2$X.
 In these papers,~\cite{rf:6,rf:9,rf:10}
however, only warping of the Fermi surface in $k_z$ direction normal 
to $H$ has been considered, and
that in $k_y$ direction is neglected.

As $t_b$ increases, 
quasi-one dimensional system turns into quasi-two dimensional 
system. Quasi-two-dimensional (Q2D) electrons, 
for example, $\beta$-(BEDT-TTF)$
\mbox{}_2$I$\mbox{}_3$ ($t_a\sim 3000$K, $t_b/t_a\sim 0.5$ and 
$t_c/t_a\sim 0.02$) have a two-dimensional cylindrical 
Fermi surface with a weak warping along $k_z$ direction.
When the magnetic field $H$ is applied in $b$ direction,
 the electrons move either in open orbit reaching the zone boundary
or in closed orbit as is schematically shown in Fig.1. \\
\begin{figure}
\begin{center}
\leavevmode
\epsfxsize=4.5cm
\epsfbox{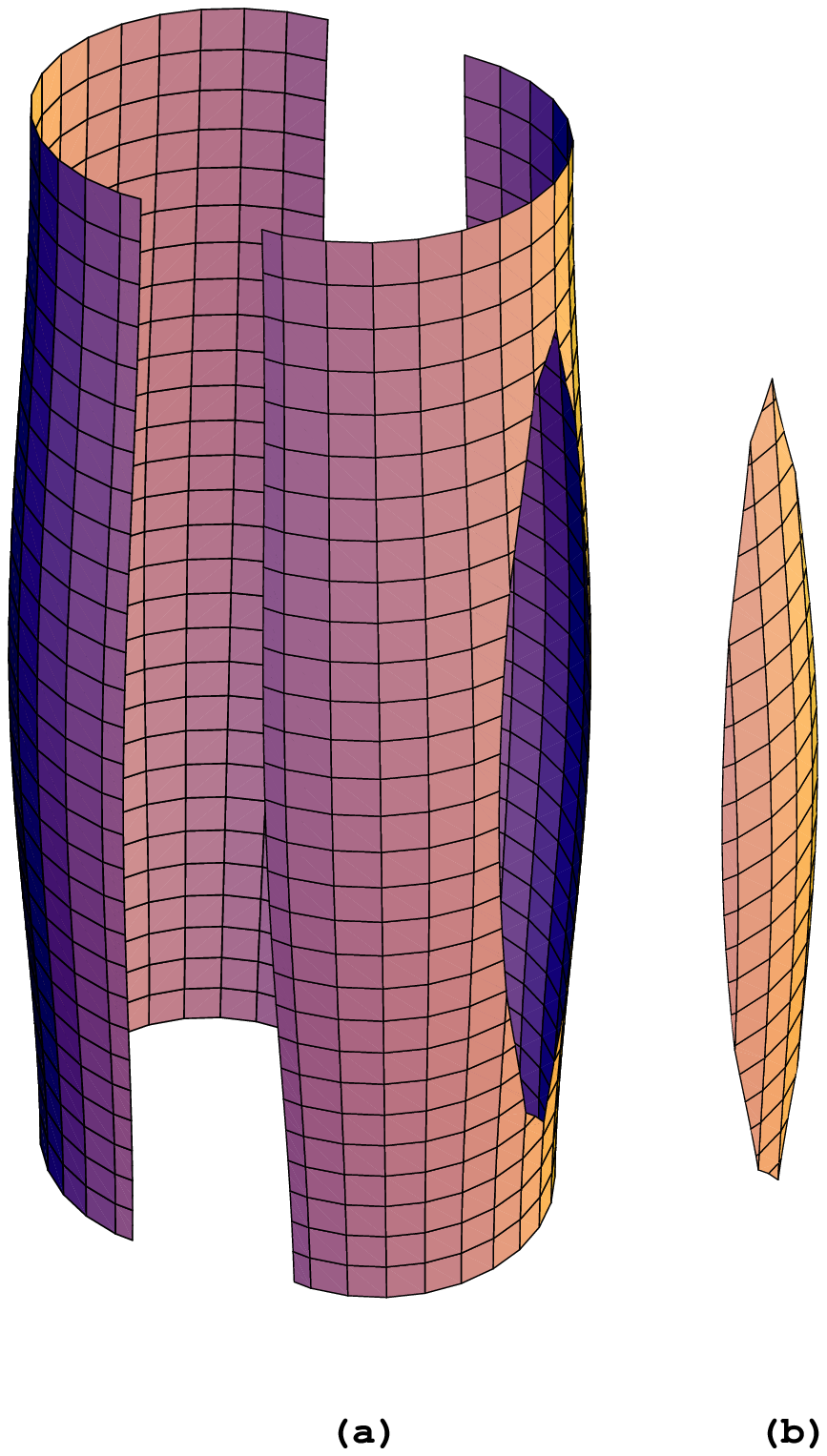}
\end{center}
\caption{Fermi surface of a quasi-two dimensional conductor is drawn. In
the presence of magnetic field along $k_y$, the electrons move open
orbit (a) and closed orbit (b).}
\label{fig1}
\end{figure}\\
The former gives the similar effect as in Q1D superconductors.
It have been discussed by Lebed~\cite{rf:6} and Dupuis {\it et al}.
~\cite{rf:9}
 that the Q2D superconductors will evolve from the GLAG region to
reentrant phase.
Recently, Lebed and Yamaji~\cite{rf:11} have calculated the mean
field transition temperature of Q2D superconductor and
 shown the reentrant behavior. They have used the parabolic band in the
conducting plane, i.e. the lattice structure is neglected in the plane.

In this paper, we calculate the mean field transition temperature 
numerically by taking account of the eigenstates of three-dimensional 
tight-binding electrons in magnetic field. 
In this formation, we can treat
both Q1D and Q2D systems by changing $t_b/t_a$.

We start from the tight-binding Hamiltonian 
(we take $\hbar=k_{\rm B}=c_0=1$, where $c_0$ is a light velocity, 
throughout the paper),
\begin{eqnarray}
{\cal H}_0&=&-t_a\sum_{(i,j)_a,\sigma}e^{{\rm i}\theta_{ij}}
c^{\dag}_{i,\sigma}c_{j,\sigma}-t_b\sum_{(i,j)_b,\sigma}
e^{{\rm i}\theta_{ij}}c^{\dag}_{i,\sigma}c_{j,\sigma}
\nonumber\\
& &-t_c\sum_{(i,j)_c,\sigma}e^{{\rm i}\theta_{ij}}c^{\dag}_{i,\sigma}
c_{j,\sigma}
-\mu\sum_{i,\sigma}c^{\dag}_{i,\sigma}c_{i,\sigma}\nonumber\\
& &-\sum_{i,\sigma}\sigma\mu_{\rm B}Hc^{\dag}_{i,\sigma}c_{i,\sigma},
\end{eqnarray}
where $c^{\dag}_{i,\sigma}$ and $c_{i,\sigma}$ are creation and
annihilation operators, $\mu$ is chemical 
potential and $\sigma\mu_{\rm B}H$ is the Zeeman energy for $\uparrow$ 
($\downarrow$) spin ($\sigma=+(-)$) and 
\begin{equation}
\theta_{ij}=\frac{2\pi}{\phi_0}\int^j_i\mbox{\boldmath
$A$}d\mbox{\boldmath$l$}.
\end{equation}

In the above ${\mib A}$ is the vector potential. 
We consider the anisotropic system with 
the hopping matrix elements $t_a\geq t_b\gg t_c$.
 We neglect the field dependence of $\mu$, 
since the energy gap due to the magnetic field is very small in the
anisotropic system with $\phi/\phi_0\ll 1$ except for the bottom and the
top of the band. 

In this paper the magnetic field $H$ is applied in the $b$ direction.
 We take the vector potential
${\mib A}$ as ${\mib A}=(0,0,-Hx)$ and the non-interacting
Hamiltonian is written as 
\begin{eqnarray}
& &{\cal H}_0=\nonumber\\
& &\sum_{\sigma,\mbox{\scriptsize{\mib k}}}
C^{\dag}_{\sigma}
\left(\begin{array}{ccc}\ddots&&V^*\\&\begin{array}{ccc}M_{-1}&V&0\\V^*&M_0&
V\\0&V^*&M_1\end{array}&\\V&&\ddots\end{array}\right)C_{\sigma},
\end{eqnarray}
where 
\begin{equation}
M_{n}=-2t_a\cos[a(k_x+nG)]-2t_b\cos(bk_y)-\mu-\sigma\mu_{\rm B}H,
\end{equation}
\begin{equation}
V=-t_ce^{{\rm i}ck_z},
\end{equation}
\begin{equation}
C^{\dag}_{\sigma}=(\cdots,c^{\dag}_{\sigma}({\mib k}-{\mib
G}),c^{\dag}_{\sigma}
({\mib k}),c^{\dag}_{\sigma}({\mib k}+{\mib G}),\cdots),
\end{equation}
\begin{equation}
{\mib G}=(G,0,0)=\left(\frac{2\pi}{a}\frac{\phi}{\phi_0},0,0\right). 
\end{equation}

The creation operators of electrons can be written in terms of the
creation operators of the eigenstates 
($\Psi^{\dag}_{\sigma}(n,{\mib k})$) of eq.(3) as
\begin{equation}
c^{\dag}_{\sigma}({\mib k}+m{\mib G})=e^{{\rm i}mck_z}\sum_n
\phi^*_{k_x,k_z}(m,n)\Psi^{\dag}_{\sigma}(n,{\mib k}),
\end{equation}
where $m$ and $n$ are integers.

Using eq.(8),
the real space one-particle Green's function is given by
\begin{eqnarray}
G_{\sigma}({\mib r},{\mib r}',{\rm i}\omega_l)&=&-\int^{\beta}_0d\tau
e^{{\rm i}\omega_l\tau}\left<T_{\tau}C_{{\mib r},\sigma}(\tau)
C^{\dag}_{{\mib r}',\sigma}(0)\right>\nonumber\\
&=&\sum_{{\mib k},n}\sum_{m,m'}\frac{\phi_{k_x,k_z}
(m,n)\phi^*_{k_x,k_z}(m',n)}
{{\rm i}\omega_l-\varepsilon_{n,{\mib k},\sigma}}\nonumber\\
&\times& e^{{\rm i}(m'-m)ck_z}e^{{\rm i}({\mib r}'-{\mib r})\cdot
{\mib k}+{\rm i}(m'{\mib r}'-m{\mib r})\cdot{\mib G}},
\end{eqnarray} 
where $\omega_l=(2l+1)\pi T$ is a Matsubara frequency and $l$ is integer. 
In this paper, we neglect the Zeeman energy for simplicity. 
The coefficients $\phi_{k_x,k_z}(m,n)$ and eigenvalues
\begin{equation}
\varepsilon_{n,{\mib k}}=\epsilon(n,k_x,k_z)-2t_b\cos(bk_y)-\mu, 
\end{equation}
can be calculated 
by diagonalizing the matrix in eq.(3) numerically, where
$\epsilon(n,k_x,k_z)$ is the eigenvalue of eq.(3) for 
$-2t_b\cos(bk_y)-\mu=0$. If $\phi/\phi_0=p/q$,
 where $p$ and $q$ are mutually prime integers, the matrix size of eq.(3) 
is $q\times q$ and the magnetic Brillouin zone is given by 
$-\pi/(qa)<k_x<\pi/(qa)$.

Since we are interested in the instability to the superconductivity in
the quarter-filled tight-binding electrons at the temperature much
smaller than the band width, the eigenstates near the top of the band
are not important in the anisotropic hopping case ($t_c\ll t_a$). Thus,
we calculate only $3/4\sim 1/2$ of the states from the bottom of 
the band in eq.(3) by neglecting $V$ and $V^*$ for $c(\mib k+n\mib G)$
 with $|k_x+nG|\geq (3/4)\pi/a$ or $|k_x+nG|\geq (1/2)\pi/a$. We have 
checked that the result is not changed by this approximation. 
In this approximation, we can take the matrix size as 
$\sim[\pi/G]\times[\pi/G]$ and $-G/2<k_x<G/2$. 
The eigenvalues and coefficients do not depend on $k_z$ in this 
approximation and we write $\epsilon(n,k_x,k_z)=\epsilon(n,k_x)$ and 
$\phi_{k_x,k_z}(m,n)=\phi_{k_x}(m,n)$ ($\phi_{k_x}(m,n)$ can be taken
real).

In the mean field approximation, the linearized gap equation for 
s-wave pairing in coordinate representation is obtained as 
\begin{equation}
\Delta({\mib r})=\lambda\int d{\mib r}'
K({\mib r},{\mib r}')\Delta({\mib r}'), 
\end{equation}
where
\begin{equation}
K({\mib r},{\mib r}')\equiv T\sum_{\omega_l}G
({\mib r},{\mib r}',{\rm i}\omega_l)G
({\mib r},{\mib r}',-{\rm i}\omega_l),
\end{equation}
and $\lambda$ is the coupling constant.
We write 
\begin{equation}
\Delta(x)=\sum_{q_x,N}e^{{\rm i}(q_x+NG)x}\Delta_{N}(q_x),
\end{equation}
where $q_x$ is taken as $-G/2<q_x\le G/2$.
For even $N$, using eqs.(9) and (13),  
linearized gap equation is written as a matrix equation 
\begin{equation}
\Delta_{2j}(q_x)=\lambda\sum_{j'}\Pi_{2j,2j'}\Delta_{2j'}(q_x),
\end{equation}
where
\begin{eqnarray}
& &\Pi_{2j,2j'}(q_x)=\sum_{k_x,k_y}\sum_{n,n'}\sum_m\phi_{k_x}(m-j,n)
\phi_{k_x}(m-j',n)
\nonumber\\
& &\times \phi_{-k_x}(-m-j,n')\phi_{-k_x}(-m-j',n')
\nonumber\\
& &\times \frac{1-f(\varepsilon_{n,k_x,k_y})
-f(\varepsilon_{n',-k_x-q_x,-k_y})}
{2(\varepsilon_{n,k_x,k_y}+\varepsilon_{n',-k_x-q_x,-k_y})},
\end{eqnarray}
where $f(\varepsilon)$ is the Fermi distribution function.

For odd $N$, we get
\begin{equation}
\Delta_{2j+1}(q_x)=\lambda\sum_{j'}\Pi_{2j+1,2j'+1}\Delta_{2j'+1}(q_x),
\end{equation}
where
\begin{eqnarray}
& &\Pi_{2j+1,2j'+1}(q_x)=\sum_{k_x,k_y}\sum_{n,n'}\sum_m\phi_{k_x}(m-j,n)
\phi_{k_x}(m-j',n)
\nonumber\\
& &\times \phi_{-k_x}(-m-j-1,n')\phi_{-k_x}(-m-j'-1,n')
\nonumber\\
& &\times \frac{1-f(\varepsilon_{n,k_x,k_y})
-f(\varepsilon_{n',-k_x-q_x,-k_y})}{2(\varepsilon_{n,k_x,k_y}
+\varepsilon_{n',-k_x-q_x,-k_y})}.
\end{eqnarray}

The transition line is given by $1-g\lambda=0$ for eq.(14) and eq.(16),
where $g$ is the maximum eigenvalue of the matrix $\Pi$ of the even part
or the odd part.
In this paper, we calculate the field dependence of the 
effective coupling constant $g$ at low
temperature instead of calculating the transition temperature. 
In the BCS theory Cooper pairs are formed by electrons with wave numbers
$\mib k$ and $-\mib k$ and the energy of these states are different in
the presence of magnetic field, which causes the orbital frustration. In
the formulation of eqs.(14) and (16), however, we can take the states
($n,\mib k$) and ($n',-\mib k$) which have the same energy
$\varepsilon_{n,k_x,k_y}=\varepsilon_{n',-k_x,-k_y}$. Therefore the
superconductivity is not destroyed by the orbital frustration in the
strong magnetic field. This is the similar mechanism for the
FISDW.~\cite{rf:17} In the weak magnetic field the coefficient
$\phi_{k_x}(m,n)$ becomes small and the present results reproduce 
the GLAG results.

For the Q1D superconductors with open Fermi surface,  
the energy dispersion in $k_x$ is taken to be linear and the Fermi 
velocity
and the density of states are taken to be constant in the previous
calculations.~\cite{rf:6,rf:9,rf:10} As a first 
approximation, we take 
\begin{equation}
M_n\approx {\rm sgn}(n)v_{\rm F}(k_y)(|k_x+nG|-k_{\rm F}),
\end{equation}
where $v_F(k_y)=2t_aa\sin ak_{\rm F}(k_y)$ and $k_F(k_y)$ are the Fermi 
velocity and Fermi wave number depending on $k_y$, respectively. We may
diagonalize the matrix for $k_x+nG\sim k_{\rm F}(k_y)$ and $k_x+nG\sim
-k_{\rm F}(k_y)$ independently, as in the previous calculations. Then
the eigenstates are given with the coefficients,
\begin{equation}
\phi_{k_x}(m,n)\cong J_{m-n}
\left(\frac{2t_c}{v_{\rm F}(k_y)G}\right)\
\ {\rm for}\ n,m>0
\end{equation}
and
\begin{equation}
\phi_{k_x}(m,n)\cong J_{-m+n}
\left(\frac{2t_c}{v_{\rm F}(k_y)G}\right)\ 
\ {\rm for}\ n,m<0,
\end{equation}
where $J$ is the Bessel function. Note that in this approximation 
$\phi_{k_x}(m,n)$ does not depend on $k_x$ but it depends on $k_y$
through $v_{\rm F}(k_y)$.
In Fig.2, we plot the effective coupling constant $g/g_0$ obtained by
this linearized dispersion by solid, dashed, and dot-dashed lines as a 
function of $aG/(2\pi)$, where $g_0$ is the effective coupling 
constant for $t_c=0$ 
, which corresponds to that in the absence of magnetic field. \\
\begin{figure}
\begin{center}
\leavevmode
\epsfxsize=6.5cm
\epsfbox{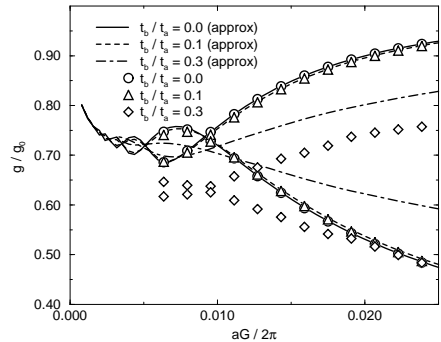}
\end{center}
\caption{Effective coupling constant as a function of $aG/(2\pi)$
 in the case of $t_c/t_a=0.02$ and $T/t_a=0.001$. In strong magnetic
field, the effective coupling
constant obtained by diagonalizing the even part of the matrix $\Pi$
reaches to that of zero magnetic field. The coupling constant for odd
part gives zero in strong magnetic field. Solid, dashed and dot-dashed
lines are obtained in the approximation of linearized energy
dispersion. Circle, triangle and diamond symbols are obtained without
that approximation.}
\label{fig2}
\end{figure}\\
In strong magnetic field, the effective coupling constant
of the even part is increased and that of the odd part is decreased for
each $t_b/t_a$.
In the case of $t_b=0$, we get the previous result.~\cite{rf:9} 
As can be seen in Fig.2, for larger $t_b/t_a$ the oscillation becomes 
small.

Next, we calculate the effective coupling
constant $g/g_0$ by numerically diagonalizing the lower $3/4$ or $1/2$ 
of the matrix in eq.(3) without using the approximation eq.(18) and we 
plot the results as circles, triangles and diamonds in Fig.2. For
$t_b/t_a=0.1$ the results are almost same as that obtained by the 
approximation with the $k_y$-dependent Fermi velocity (eq.(18)) as
expected, but the deviation is large for larger $t_b/t_a$.  

We also study the quasi-two dimensional superconductor with
$t_b/t_a=0.5$ and $1.0$. 
In Fig.3 we plot the effective coupling constant as a function of 
$aG/(2\pi)$. 
 As is seen in Fig.3, the effective coupling constant reaches to that 
for $t_c=0$ as magnetic field is increased. \\
\begin{figure}
\begin{center}
\leavevmode
\epsfxsize=6.5cm
\epsfbox{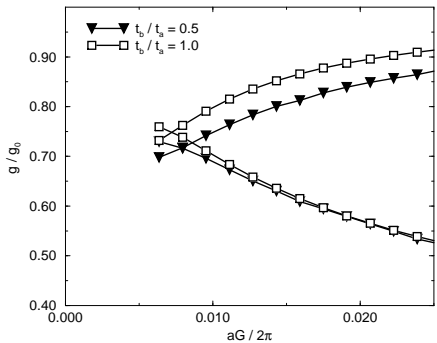}
\end{center}
\caption{Effective coupling constant as a function of $aG/(2\pi)$
 in the case of $t_c/t_a=0.02$ and $T/t_a=0.001$.}
\label{fig3}
\end{figure}\\
The reason why $g/g_0$ is small for $t_b/t_a=0.3$ is as follows. 
In Fig.4, we plot the effective coupling constant as a function of 
$t_b/t_a$. In the case of $t_c=0$, there is a 
logarithmic divergence at about $t_b/t_a\sim 0.3$, 
which is van Hove singularity for the quarter-filled band.\\
\begin{figure}
\begin{center}
\leavevmode
\epsfxsize=6.5cm
\epsfbox{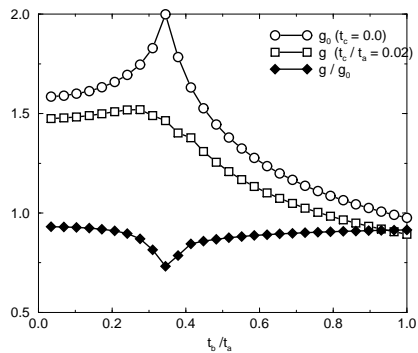}
\end{center}
\caption{Effective coupling constant as a function of $t_b/t_a$ at
 $aG/(2\pi)$=0.025 and $T/t_a=0.001$.}
\label{fig4}
\end{figure}\\
Therefore, the effective coupling constant normalizes by that for 
$t_c=0$ is small for $t_b/t_a\sim 0.3$.

In this paper we have neglected the Zeeman term for simplicity. 
However, we can
calculate the transition line with Zeeman term in this
expression. The Zeeman term
does not play any important role for the equal-spin-pairing case of the
spin triplet. If the Zeeman energy is taken into account, the transition
temperature of spin singlet is reduced due to the effect of Pauli
pair-breaking except for the Q1D case ($t_b=0$). The superconductivity of
Q1D systems is not
completely destroyed,~\cite{rf:6,rf:9,rf:10} since half of the density of 
states is available 
to make Cooper pairs for the Larkin-Ovchinnikov-Fulde-Ferrell (LOFF)
state,~\cite{rf:18,rf:19} as in the pure 1D
case.~\cite{rf:20,rf:21,rf:22}

In conclusion, we have shown the transition lines of
quasi-one dimensional and quasi-two dimensional superconductors
in tight-binding model. 
The Green function in Q1D systems is described by the Bessel 
function if we apply the
approximation that the 
energy dispersion in $k_x$ direction is taken to be linear.
In this paper, we have shown that the Green
function of the tight-binding electrons is numerically calculated without
using the linearization of the energy dispersion. In the strong
magnetic field Cooper pairs are formed in the eigenstates with the same
energy. We have
obtained the transition line $T_c(H)$ for both Q1D and Q2D cases.
As $H$ becomes large, $T_c(H)$ increases oscillationally in both cases.

\end{document}